\documentclass{article}
\usepackage{amsmath, amssymb, bbm}
\usepackage{cite}

\begin{document}
\title{On the physical interpretation of the Dirac wavefunction}
\author{Anastasios Y. Papaioannou \\ \tt{tasosp@gmail.com}}

\maketitle

\begin{abstract}
Using the language of the Geometric Algebra, we recast the massless Dirac bispinor as a set of Lorentz scalar, bivector, and pseudoscalar fields that obey a generalized form of Maxwell's equations of electromagnetism. The spinor's unusual $4 \pi$ rotation symmetry is seen to be a mathematical artifact of the projection of these fields onto an abstract vector space, and not a physical property of the dynamical fields themselves. We also find a deeper understanding of the spin angular momentum and other Dirac field bilinears in terms of these fields and their corresponding analogues in classical electromagnetism.

\end{abstract}

\section{Introduction}
\label{sec:introduction}

The interpretation of the wavefunction remains a mystery at the heart of quantum theory. While the practical success of quantum mechanics and quantum field theory is undeniable, the Dirac bispinor, and the matrix-valued algebra which acts upon it, lacks a clear physical interpretation. In the present work, we use the Geometric Algebra to point the way towards a clearer understanding of the physical structure and transformation properties of the Dirac bispinor.

We begin in section \ref{sec:summary} by summarizing the Geometric Algebra as it applies to $(3+1)$-dimensional relativistic physics. Section \ref{sec:massless} then applies these concepts to the massless Dirac equation, and develops a formulation of the Dirac bispinor in which the dynamical and spinorial degrees of freedom are distinct. With this separation of the degrees of freedom, the Dirac equation takes the form of Maxwell's equations of classical electromagnetism, with electric and magnetic current densities. Section \ref{sec:bilinears} then constructs the Dirac field bilinears, with special emphasis on their parallels with the bilinears of classical electromagnetism. Finally, in section \ref{sec:lorentz} we analyze the Lorentz transformation properties of the fields, showing how the dynamic and spinorial degrees of freedom transform separately.

\section{Summary of the Geometric Algebra}
\label{sec:summary}

In this section, we briefly summarize the Geometric Algebra and its application to relativistic physics. For a much deeper mathematical development, and a broader range of applications in classical and modern physics, we refer the reader to works by Hestenes \cite{hestenes-space-time-algebra}, \cite{hestenes-new-foundations} and Doran and Lasenby \cite{doran-lasenby}.

The algebra is constructed from four basis vector elements $e_\mu \, (\mu = 0, 1, 2, 3)$, subject to the anticommutation law
\begin{equation}
e_\mu \cdot e_\nu \equiv \frac{1}{2} (e_\mu e_\nu + e_\nu e_\mu) = g_{\mu \nu} \, 1,
\end{equation}
where $1$ is the scalar identity element of the algebra. We use the metric with signature $g_{\mu \nu} = \text{diag(+ - - -)}$, so that $e_0^2 = - e_1^2 = - e_2^2= - e_2^2 = 1$. The anticommutation law defines the symmetric \textit{inner product} $e_\mu \cdot e_\nu$ of two vectors and directly implies that any two orthogonal vectors anticommute. The antisymmetric \textit{wedge} or \textit{outer product} $e_\mu \wedge e_\nu$ of two vectors is:
\begin{equation}
e_\mu \wedge e_\nu \equiv \frac{1}{2} (e_\mu e_\nu - e_\nu e_\mu) = - (e_\nu \wedge e_\mu).
\end{equation}
Aside from the new terminology and notation for the inner and outer product, these operations are familiar from the Dirac algebra and its associated matrices $\{ \gamma_\mu \}$. We emphasize, however, that although the algebraic structure is the same, we are not working within a (matrix) representation, but with algebraic elements $\{ e_\mu \}$ interpreted as unit basis vectors. The outer product $e_\mu \wedge e_\nu$ is a new two-dimensional, or \textit{grade}-2, element of the algebra, the \textit{bivector}. (The scalar and vector elements are grade-0 and grade-1, respectively.) The full associative, non-commutative product of two vectors is the sum of their inner and outer products:
\begin{equation}
e_\mu e_\nu = e_\mu \cdot e_\nu + e_\mu \wedge e_\nu.
\end{equation}

The product of two vectors $v = v^\mu e_\mu$ and $w = w^\nu e_\nu$ will in general contain both scalar and bivector terms:
\begin{align*}
v w & = (v^\mu e_\mu) (w^\nu e_\nu) = (v^\mu w^\nu) (e_\mu e_\nu ) \\
& = v^\mu w^\nu (e_\mu \cdot e_\nu + e_\mu \wedge e_\nu) \\
& = v^\mu w^\nu (e_\mu \cdot e_\nu) + v^\mu w^\nu (e_\mu \wedge e_\nu) \\
& = v \cdot w + v \wedge w.
\end{align*}
The inner product is the familiar scalar product of two vectors:
\begin{align*}
v \cdot w & = v^\mu w^\nu (e_\mu \cdot e_\nu) \\
& = v^\mu w^\nu g_{\mu \nu} \, 1 \\
& = v^\mu w_\mu \, 1,
\end{align*}
while the outer product is the relativistic analogue of the vector cross product $\vec{v} \times \vec{w}$:
\begin{align*}
v^\mu w^\nu (e_\mu \wedge e_\nu) & = v^\nu w^\mu (e_\nu \wedge e_\mu) \\
& = -v^\nu w^\mu (e_\mu \wedge e_\nu) \\
& = \frac{1}{2} (v^\mu w^\nu - v^\nu w^\mu) (e_\mu \wedge e_\nu).
\end{align*}
Unlike, however, the vector cross product $e_i \times e_j = \epsilon_{i j k} e_k$, which multiplies two vectors to form a third (axial) vector, the bivector $e_\mu \wedge e_\nu$ is a new two-dimensional element of the algebra, which is better viewed not as a vector but as an oriented area in the plane defined by $e_\mu$ and $e_\nu$.

Six linearly independent bivectors can be constructed from the basis vectors:
\begin{align*}
e_1 \wedge e_0 & \qquad e_2 \wedge e_0 \\
e_3 \wedge e_0 & \qquad e_1 \wedge e_2 \\
e_2 \wedge e_3 & \qquad e_3 \wedge e_1.
\end{align*}
(We have explicitly written these bivectors using the wedge operator to indicate that the antisymmetric outer product has been used. When two vectors $v$ and $w$ are orthogonal, the outer product $v \wedge w$ is equivalent to the full product $v w$. In cases where this orthogonality is clear, e.g., for a timelike bivector $e_i \wedge e_0$, we will often simplify our notation and write the product as $e_i e_0$.)

The bivectors inherit their multiplicative properties from the vectors. For example, the square of the bivector $e_1 \wedge e_0$ is
\begin{align*}
(e_1 \wedge e_0)^2 & = (e_1 e_0)(e_1 e_0) \\
& = e_1 (e_0 e_1) e_0 \\
& = - e_1 (e_1 e_0) e_0 \\
& = - (e_1 e_1) (e_0 e_0) \\
& = - g_{1 1} \, g_{0 0} \\
& = +1.
\end{align*}
Similar calculations for the other bivector basis elements show that the ``timelike'' bivectors of the form $e_i \wedge e_0$ all square to $+1$, while the ``spacelike'' bivectors $e_i \wedge e_j$ square to $-1$.

Likewise, the multiplication of vectors with bivectors will also follow from the multiplicative properties of the constituent vectors. The product of $e_1$ with $e_1 \wedge e_2$, for example, is the vector $-e_2$:
\begin{align*}
e_1 (e_1 \wedge e_2) & = e_1 (e_1 e_2) \\
& = (e_1 e_1) e_2 \\
& = g_{1 1} \, e_2 \\
& = - e_2,
\end{align*}
while the product of $e_2$ with the same bivector is the vector $e_1$:
\begin{align*}
e_2 (e_1 \wedge e_2) & = e_2 (e_1 e_2) \\
& = - e_2 (e_2 e_1) \\
& = - (e_2 e_2) e_1 \\
& = - g_{2 2} \, e_1 \\
& = e_1.
\end{align*}
Multiplication by $e_1 \wedge e_2$ has thus rotated the elements clockwise by $\pi/2$ in the $e_1 - e_2$ plane:
\begin{align*}
e_1 & \to -e_2, \\
e_2 & \to e_1.
\end{align*}
We will make further use of this property of bivectors when analyzing Lorentz transformations.

The product of $e_3$ with $e_1 \wedge e_2$, however, is not a vector, but a new grade-3 element, a \textit{pseudovector}:
\begin{align*}
e_3 (e_1 \wedge e_2) & = e_3 (e_1 e_2) \\
& = e_1 e_2 e_3 \\
& \equiv e_1 \wedge e_2 \wedge e_3.
\end{align*}
Here we have extended the outer product to higher grades: for vectors $v_1$, $v_2$, and $v_3$, $v_1 \wedge v_2 \wedge v_3$ is the grade-3 term(s) of the product $v_1 v_2 v_3$. More generally, the outer product of $n$ vectors, $v_1 \wedge v_2 \wedge \cdots \wedge v_n$ will be the grade-$n$ term(s) of the product $v_1 v_2 \cdots v_n$. By the anticommutation properties of its constituent vectors, the outer product is fully antisymmetric upon interchange of any pair of vector indices. This full antisymmetry also guarantees that the outer product is associative, e.g., $v_1 \wedge (v_2 \wedge v_3) = (v_1 \wedge v_2) \wedge v_3 = v_1 \wedge v_2 \wedge v_3$.

There are four linearly independent pseudovectors:
\begin{align*}
e_0 \wedge e_1 \wedge e_2 & \qquad e_0 \wedge e_2 \wedge e_3 \\
e_0 \wedge e_3 \wedge e_1 & \qquad e_1 \wedge e_2 \wedge e_3.
\end{align*}
The three pseudovectors containing $e_0$ all square to $-1$, e.g.,
\begin{align*}
(e_0 \wedge e_1 \wedge e_2)^2 & = (e_0 e_1 e_2) (e_0 e_1 e_2) \\
& = - e_0 e_0 e_1 e_1 e_2 e_2 \\
& = - g_{0 0} \, g_{1 1} \, g_{2 2} \\
& = - 1,
\end{align*}
while the remaining pseudovector $e_1 \wedge e_2 \wedge e_3$ squares to $+1$:
\begin{align*}
(e_1 \wedge e_2 \wedge e_3)^2 & = (e_1 e_2 e_3)(e_1 e_2 e_3) \\
& = - e_1 e_1 e_2 e_2 e_3 e_3 \\
& = g_{1 1} \, g_{2 2} \, g_{3 3} \\
& = +1.
\end{align*}

There is, finally, one unique grade-4 \textit{pseudoscalar},
\begin{equation*}
I \equiv e_0 \wedge e_1 \wedge e_2 \wedge e_3,
\end{equation*}
which squares to $-1$:
\begin{align*}
I^2 & = (e_0 e_1 e_2 e_3) (e_0 e_1 e_2 e_3) \\
& = + (e_0 e_0) (e_1 e_1) (e_2 e_2) (e_3 e_3) \\
& = g_{0 0} \, g_{1 1} \, g_{2 2} \, g_{3 3} \\
& = -1.
\end{align*}

In total there are 16 linearly-independent basis elements of the algebra: 1 scalar, 4 vectors, 6 bivectors, 4 pseudovectors, and 1 pseudoscalar. Any vector multiplying the pseudoscalar $I$ will contract with one of its constituent basis vectors, so that no elements of grade 5 or higher are possible. The basis contains six square roots of $+1$:
\begin{gather*}
1 \\
e_0 \\
e_1 \wedge e_0 \quad e_2 \wedge e_0 \quad e_3 \wedge e_0 \\
e_1 \wedge e_2 \wedge e_3,
\end{gather*}
and ten roots of $-1$:
\begin{gather*}
e_1 \quad e_2 \quad e_3 \\
e_1 \wedge e_2 \quad e_2 \wedge e_3 \quad e_3 \wedge e_1 \\
e_0 \wedge e_1 \wedge e_2 \quad e_0 \wedge e_2 \wedge e_3 \quad e_0 \wedge e_3 \wedge e_1 \\
I.
\end{gather*}

\subsection{Grade Parity, Duality, and the Multivector}
We call \textit{even} those quantities that have an even grade, namely the scalar, bivectors, and pseudoscalars; those with odd grade (the vectors and pseudovectors) are called \textit{odd}. This concept of ``grade parity'' will be extremely useful when analyzing bilinear quantities: any product containing only even quantities (or an even number of odd quantities) will necessarily also be even, and any product containing an odd number of odd quantities will necessarily be odd. This allows us to analyze complicated products and to determine immediately, based on the grade parity, which terms exist and which are necessarily zero.

In later sections we will make extensive use of a few important properties of $I$. As previously noted, it is a root of $-1$. Also, because it contains one factor of each basis vector, it anticommutes all odd elements and commutes with all even elements (including itself). Finally, $I$ serves as a duality operator, both between the spacelike and timelike bivectors:
\begin{equation*}
e_i \wedge e_j = - \epsilon_{i j k} I e_k e_0,
\end{equation*}
and between the vectors and pseudovectors:
\begin{align*}
e_1 \wedge e_2 \wedge e_3 & = - I e_0 \\
e_0 \wedge e_i \wedge e_j & = -\epsilon_{i j k} I e_k.
\end{align*}
We will use this duality operation to greatly simplify our calculations below.

A general element of the algebra, the \textit{multivector}, is a sum of scalar, bivector, pseudovector, and pseudoscalar terms:
\begin{equation*}
M = f \, 1 + v^\mu e_\mu + \frac{1}{2!} F^{\mu \nu} e_\mu \wedge e_\nu + \frac{1}{3!} p^{\mu \nu \rho} e_\mu \wedge e_\nu \wedge e_\rho + \frac{1}{4!} g^{\mu \nu \rho \sigma} e_\mu \wedge e_\nu \wedge e_\rho \wedge e_\sigma.
\end{equation*}
By analogy with the bivector field $F$ of electromagnetism, we can write $F^{i 0} = - F_{i 0} = E_i$ and $F^{i j} = F_{i j} = - \epsilon^{i j k} B_k$ for the bivector field components, and simplify the pseudovector and pseudoscalar field components:
\begin{align*}
p^\mu I e_\mu & \equiv \frac{1}{3!} p^{\mu \nu \rho} e_\mu \wedge e_\nu \wedge e_\rho \\
g \, I & \equiv \frac{1}{4!} g^{\mu \nu \rho \sigma} e_\mu \wedge e_\nu \wedge e_\rho \wedge e_\sigma.
\end{align*}
With these changes, the multivector takes the form
\begin{equation}
\psi_M = f \, 1 + v^\mu e_\mu + E_i \, e_i e_0 + B_i \, I e_i e_0 + p^\mu I e_\mu + g \, I.
\end{equation}

In the Geometric Algebra, the coefficients $f$, $v^\mu$, $E_i$, $B_i$, $p^\mu$, and $g$ take only real values. As we will see below, any imaginary values will be interpreted in terms of the basis elements that are roots of $-1$.

\subsection{Grade Extraction, the Hermitian Adjoint, and Reversal}

For a general multivector $M$, we will often wish to isolate terms of a particular grade. $\langle M \rangle_r$ denotes the term(s) with grade $r$. If the subscript is omitted, the scalar (grade-0) term is implied: $\langle M \rangle \equiv \langle M \rangle_0$. One property of the scalar term that will prove extremely useful is that a product of multivectors is unaffected by cyclic permutations:
\begin{equation*}
\langle M_1 M_2 \cdots M_n \rangle = \langle M_n M_1 M_2 \cdots M_{n-1} \rangle.
\end{equation*}

As mentioned previously, we can generalize the inner and outer products between vectors to higher-grade objects. For multivectors $A_r$ and $B_s$ with respective non-zero grades $r$ and $s$, the product $A_r B_s$ will contain terms with grades $|r-s|$, $|r-s|+2$, etc., up to $r+s$. The inner product denotes the term(s) with minimum grade $|r-s|$:
\begin{equation*}
A_r \cdot B_s \equiv \langle A_r B_s \rangle_{|r-s|},
\end{equation*}
and the outer product $A_r \wedge B_s$ will be defined as the term(s) in the product $A_r B_s$ with the maximum grade $r+s$:
\begin{equation*}
A_r \wedge B_s \equiv \langle A_r B_s \rangle_{r+s}.
\end{equation*}

The Hermitian adjoint of a product of vectors is the reverse-ordered product of the Hermitian adjoint of the individual vectors:
\begin{equation*}
( e_{\mu_1} e_{\mu_2} \cdots e_{\mu_n} )^\dagger = {e_{\mu_n}}^\dagger \cdots {e_{\mu_2}}^\dagger {e_{\mu_1}}^\dagger
\end{equation*}
In both the Weyl and Dirac bases, the Hermitian adjoint changes the sign of $\gamma^i$, leaving $\gamma^0$ unchanged:
\begin{align*}
{\gamma^0}^\dagger &= \gamma^0 \\
{\gamma^i}^\dagger &= -\gamma^i.
\end{align*}
The corresponding operation in the Geometric Algebra ($e_0 \to e_0$, $e_i \to -e_i$) is equivalent to a spatial reflection through the origin. This operation can be written as a two-sided transformation:
\begin{equation*}
e_\mu \to e_\mu^\dagger = e_0 e_\mu e_0.
\end{equation*}
The Hermitian adjoint of a product of basis vectors then becomes:
\begin{align}
( e_{\mu_1} e_{\mu_2} \cdots e_{\mu_n} )^\dagger \nonumber
&= {e_{\mu_n}}^\dagger \cdots {e_{\mu_2}}^\dagger {e_{\mu_1}}^\dagger \nonumber \\
&= ( e_0 e_{\mu_n} e_0 ) \cdots ( e_0 e_{\mu_2} e_0 ) ( e_0 e_{\mu_1} e_0 ) \nonumber \\
&= e_0 ( e_{\mu_n} \cdots \, e_{\mu_2} e_{\mu_1} ) e_0 \nonumber \\
&= e_0 ( e_{\mu_1} e_{\mu_2} \cdots \, e_{\mu_n} )^\sim e_0
\end{align}
where the reversal operation $\widetilde{M}$ reverses the order of all constituent vectors in the multivector $M$.

The net result of Hermitian conjugation depends on two factors: parity under the reversal operation, and spatial parity. Upon reversal, the scalars, vectors, and pseudoscalars are even, whereas the bivectors and pseudovectors are odd:
\begin{align*}
1^\sim & = +1 \\
(e_\mu)^\sim & = + e_\mu \\
(e_\mu \wedge e_\nu)^\sim & = - (e_\mu \wedge e_\nu) \\
(e_\mu \wedge e_\nu \wedge e_\rho)^\sim & = - (e_\mu \wedge e_\nu \wedge e_\rho) \\
(e_\mu \wedge e_\nu \wedge e_\rho \wedge e_\sigma)^\sim & = + (e_\mu \wedge e_\nu \wedge e_\rho \wedge e_\sigma).
\end{align*}
A general multivector will therefore have mixed parity under reversal.

Under spatial reflection, elements composed of an even number of spacelike vectors, namely $1$, $I e_i e_0$, $I e_i$, and $I$, are positive; those composed of an odd number ($e_i$, $e_i e_0$, $I e_0$) are negative. If an element has the same parity under both reversal and spatial reflection, then it will be its own Hermitian conjugate. The self-conjugate basis elements are:
\begin{equation*}
1 \qquad e_0 \qquad e_i e_0 \qquad I e_0.
\end{equation*}
The basis elements that have opposite parities under reversal and spatial reflection will change sign under Hermitian conjugation:
\begin{equation*}
e_i \qquad I e_i e_0 \qquad I e_i \qquad I.
\end{equation*}
The self-conjugate elements are the roots of $+1$, and the elements that change sign under conjugation are the roots of $-1$, making Hermitian conjugation the natural extension of the complex conjugation operation to the Geometric Algebra.

\section{The massless wave equation}
\label{sec:massless}

Using the Geometric Algebra, we can write the ``square root'' of the second-order differential operator $\Box = \partial_{tt} - c^2 \boldsymbol{\nabla}^2$ in terms of the first-order, vector-valued operator $\nabla = \partial^\mu e_\mu$:
\begin{equation*}
(\partial^\mu e_\mu) (\partial^\nu e_\nu) = \partial^\mu \partial^\nu \frac{1}{2} (e_\mu e_\nu + e_\nu e_\mu) = \partial^\mu \partial^\nu (g_{\mu \nu} \, 1) = \Box \, 1.
\end{equation*}
For a scalar field $f$ subject to the massless wave equation
\begin{equation*}
\Box f = 0,
\end{equation*}
we expand the number of fields from the single scalar $f$ to a multivector with 16 field degrees of freedom:
\begin{equation*}
\psi_M = f \, 1 + v^\mu e_\mu + E_i \, e_i e_0 + B_i \, I e_i e_0 + p^\mu I e_\mu + g \, I,
\end{equation*}
which is subject to the multivector-valued first-order differential equation
\begin{equation}
\nabla \psi_M = 0.
\end{equation}
The vector derivative of the scalar $f \, 1$ is a vector-valued quantity:
\begin{equation*}
(\partial^\mu e_\mu) f \, 1 = (\partial^\mu f) \, e_\mu.
\end{equation*}
The bivector field will have vector and pseudovector derivative terms:
\begin{align*}
(\partial^\mu e_\mu) (E_i \, e_i e_0 + B_i \, I e_i e_0) = & - \partial_0 E_i \, e_i + \partial_i E_i \, e_0 + \partial^j E_i \epsilon_{i j k} I e_k \\
& + \partial_0 B_i \, I e_i - \partial_j B_i \, I e_0 + \partial^j B_i \epsilon_{i j k} e_k.
\end{align*}
The derivative of the pseudoscalar will have pseudovector terms:
\begin{align*}
\nabla g \, I &= (\partial^\mu e_\mu) g \, I \\
& = (\partial^\mu g) e_\mu I \\
& = - (\partial^\mu g) I e_\mu.
\end{align*}
The first-order differential equation $\nabla \psi_M = 0$ can therefore be separated into vector-valued and pseudovector-valued equations:
\begin{align*}
& (\partial^\mu f) e_\mu  - \partial_0 E_i \, e_i + \partial_i E_i \, e_0 + \partial^j B_i \epsilon_{i j k} e_k = 0 \\
& \partial_0 B_i \, I e_i - \partial_j B_i \, I e_0 + \partial^j E_i \epsilon_{i j k} I e_k - (\partial^\mu g) I e_\mu = 0.
\end{align*}
There is no coupling of the scalar, bivector, and pseudoscalar fields to any vector or pseudovector fields that may also be present. We will therefore assume that the vector and pseudovector fields are zero.

In terms of these components, the above equations yield:
\begin{align} \label{eqn:maxwell}
\partial_t f + \boldsymbol{\nabla} \cdot \mathbf{E} & = 0 \nonumber \\
\partial_t \mathbf{E} + \boldsymbol{\nabla} f - \boldsymbol{\nabla} \times \mathbf{B} & = 0 \nonumber \\
\partial_t \mathbf{B} + \boldsymbol{\nabla} g + \boldsymbol{\nabla} \times \mathbf{E} & = 0 \nonumber \\
\partial_t g + \boldsymbol{\nabla} \cdot \mathbf{B} & = 0.
\end{align}
That is, the first-order wave equation is equivalent to Maxwell's equations of electromagnetism, with $j_e = - \nabla f$ playing the role of electric current and $j_m = - \nabla g$ that of magnetic current. The multivector $\psi_M$ has eight components, and satisfies the first-order massless wave equation, as does the (massless) bispinor in the Dirac algebra.

\subsection{Lorentz transformations}

The Geometric Algebra provides a natural unified language for describing transformations. For example, the rotation of the vector $e_1$ by angle $\varphi$ in the $e_1 - e_2$ plane can be written
\begin{equation*}
e_1 \to R(\varphi) \, e_1 = \cos{\varphi} \, e_1 + \sin{\varphi} \, e_2,
\end{equation*}
where the multivector operator $R$ is
\begin{equation*}
R(\varphi) = \exp(\varphi \, e_1 \wedge e_2) = \cos{\varphi} \, 1 + \sin{\varphi} \, e_1 \wedge e_2.
\end{equation*}
This transformation could equivalently have been written in terms of an operator that right-multiplies the vector:
\begin{equation*}
e_1 \to e_1 \, \widetilde{R}(\varphi),
\end{equation*}
or as a two-sided operation, with half of the transformation performed by each operator:
\begin{equation}
e_1 \to R(\varphi/2) \, e_1 \, \widetilde{R}(\varphi/2),
\end{equation}
where $\widetilde{R}(\varphi) = \cos \varphi \, 1 - \sin \varphi \, e_1 \wedge e_2$ is the reverse of $R(\varphi)$.
The benefit of this two-sided operation is that all elements of the algebra transform in the same way, including higher-grade elements as well as vectors that are not in the plane of rotation:
\begin{align*}
e_{\mu_1} e_{\mu_2} \cdots e_{\mu_n} & \to (R(\varphi / 2) \, e_{\mu_1} \, \widetilde{R}(\varphi / 2)) (R(\varphi / 2) \, e_{\mu_2} \, \widetilde{R}(\varphi / 2)) \cdots (R(\varphi / 2) \, e_{\mu_n} \, \widetilde{R}(\varphi / 2)) \\
& = R(\varphi / 2) (e_{\mu_1} e_{\mu_2} \cdots e_{\mu_n}) \widetilde{R}(\varphi / 2),
\end{align*}
where we have made use of the important fact that $\widetilde{R}$ is the inverse of $R$.

Lorentz boosts can be written in the same way, as a ``rotation'' with a timelike bivector:
\begin{align*}
R(\alpha) & = \exp(+ \alpha \, e_3 \wedge e_0) = \cosh{\alpha} \, 1 + \sinh{\alpha} \, e_3 \wedge e_0 \\
e_\mu & \to R(\alpha/2) e_\mu \widetilde{R}(\alpha/2),
\end{align*}
and a general proper Lorentz transformation can be written as a product of a boost with rapidity $\alpha$ and a rotation through angle $\varphi$:
\begin{align*}
R(\varphi/2, \alpha/2) & = \exp(-\frac{\varphi}{2} n_i (I e_i e_0)) \exp(\frac{\alpha}{2} n'_k (e_k e_0)) \\
e_\mu & \to R(\varphi/2, \alpha/2) e_\mu \widetilde{R}(\varphi/2, \alpha/2),
\end{align*}
where $n$ is the unit normal to the plane of rotation and $n'$ is the direction of the boost. (The extra negative sign in the factor containing $\varphi$ appears because we have chosen to write the spacelike bivector $e_j \wedge e_k$ in the dual form $- \epsilon_{i j k} I e_k e_0$.)

\subsection{The Dirac bispinor}
We now construct a correspondence between the multivector and bispinor solutions. Each contravariant basis vector $e^\mu = g^{\mu \nu} e_\nu$ in the Geometric Algebra corresponds is represented by a matrix $\gamma^\mu$ (taking care to note changes of sign when changing covariant spatial vectors to contravariant ones). The bivector and higher-grade products all carry over directly:
\begin{align*}
1 & \to 1_{4 \times 4} \\
e^\mu & \to \gamma^\mu \\
e^\mu \wedge e^\nu & \to \frac{1}{2} [\gamma^\mu, \gamma^\nu] \\
I e^\mu = (- e^0 e^1 e^2 e^3) e^\mu & \to (- \gamma^0 \gamma^1 \gamma^2 \gamma^3) \gamma^\mu = i \gamma^5 \gamma^\mu \\
I = - e^0 e^1 e^2 e^3 & \to (-\gamma^0 \gamma^1 \gamma^2 \gamma^3 ) = i \gamma^5.
\end{align*}
A matrix-valued solution to the massless wave equation follows immediately:
\begin{align}
& \psi_M = f \, 1 - E_i \, \gamma^i \gamma^0 - B_i \, I \gamma^i \gamma^0 + g I \nonumber \\
& (\partial_\mu \gamma^\mu) \psi_M = 0.
\end{align}

In short, replacing each basis vector with its corresponding gamma matrix yields a matrix-valued solution to the massless wave equation, with scalar, bivector, and pseudoscalar degrees of freedom. The familiar Dirac bispinor solution, however, is not represented by a square matrix but by a column vector upon which the Dirac matrices act. We seek therefore a method of ``projecting'' the above matrix solution onto the vector space, by choosing a constant \textit{projection column vector} or \textit{projection bispinor} $w$, which right-multiplies the matrix solution to give the bispinor solution:
\begin{align}
\psi &= \psi_M w, \nonumber \\
w &= \begin{pmatrix} \alpha \\ \beta \end{pmatrix}.
\end{align}
The choice for the constant two-component spinors $\alpha$ and $\beta$ will be informed by the chosen basis (e.g., Dirac or Weyl). Below, we will use the Dirac basis.

To find $\alpha$ and $\beta$ we first note that the imaginary scalar $i = \sqrt{-1}$ has no obvious implementation in the Geometric Algebra, i.e., there is no linear combination
\begin{equation*}
A \, 1 + B_\mu \gamma^\mu + C_{\mu \nu} \gamma^\mu \wedge \gamma^\nu + D_\mu I \gamma^\mu + E \, I = i \, 1
\end{equation*}
with real components $A$, $B_\mu$, $C_{\mu \nu}$, $D_\mu$ and $E$, that commutes with all other elements and also squares to $-1$. The implementation of $i$ will therefore involve some basis-specific operation. To find such an operation, we note that factors of $i$ can always be shifted to multiply the projection bispinor $w$:
\begin{align*}
i (M \psi) &= i (M \psi_M w) \\
&= M \psi_M (i w) \\
&= M \psi_M \begin{pmatrix} i \alpha \\ i \beta \end{pmatrix}.
\end{align*}
We seek an operation $\mathcal{O}(i)$ that will multiply the projection spinors $\alpha$ and $\beta$ by $i$:
\begin{align}
i \psi & \to \psi_M \mathcal{O}(i) w \nonumber \\
& = \psi_M \mathcal{O}(i) \begin{pmatrix} \alpha \\ \beta \end{pmatrix} \\
& = \psi_M \begin{pmatrix} i \alpha \\ i \beta \end{pmatrix}. \nonumber
\end{align}

In the Dirac basis,
\begin{align*}
1 & = \begin{pmatrix} \mathbf{1} & \mathbf{0} \\ \mathbf{0} & \mathbf{1} \end{pmatrix} \\
\gamma^0 & = \begin{pmatrix} \mathbf{1} & \mathbf{0} \\ \mathbf{0} & -\mathbf{1} \end{pmatrix} \\
\gamma^i & = \begin{pmatrix} \mathbf{0} & \boldsymbol{\sigma}^i \\ -\boldsymbol{\sigma}^i & \mathbf{0} \end{pmatrix} \\
\gamma^i \gamma^0 & = \begin{pmatrix} \mathbf{0} & -\boldsymbol{\sigma}^i \\ -\boldsymbol{\sigma}^i & \mathbf{0} \end{pmatrix} \\
I \gamma^i \gamma^0 & = \begin{pmatrix} -i \boldsymbol{\sigma}^i & \mathbf{0} \\ \mathbf{0} & -i \boldsymbol{\sigma}^i \end{pmatrix} \\
I \gamma^0 & = \begin{pmatrix} \mathbf{0} & -i \mathbf{1} \\ i \mathbf{1} & \mathbf{0} \end{pmatrix} \\
I \gamma^i & = \begin{pmatrix} -i \boldsymbol{\sigma}^i & \, \mathbf{0} \\ \mathbf{0} & i \boldsymbol{\sigma}^i \end{pmatrix} \\
I \equiv - \gamma^0 \gamma^1 \gamma^2 \gamma^3 & = \begin{pmatrix} \mathbf{0} & i \mathbf{1} \\ i \mathbf{1} & \mathbf{0} \end{pmatrix},
\end{align*}
where $\mathbf{1}$, $\mathbf{0}$, and $\boldsymbol{\sigma}^i$ are the $2 \times 2$ identity, zero, and Pauli matrices. Of these matrices, only $1$, $\gamma^0$, $I \gamma^i$, and $I \gamma^i \gamma^0$ are block diagonal, and of these only $I \gamma^1$, $I \gamma^3$, $I \gamma^1 \gamma^0$, and $I \gamma^3 \gamma^0$ also have imaginary entries. We choose $\mathcal{O}(i) = - I \gamma^3 \gamma^0 = - \gamma^1 \gamma^2$, so that
\begin{equation}
- I \gamma^3 \gamma^0 w = \begin{pmatrix} i \boldsymbol{\sigma}^3 & \mathbf{0} \\ \mathbf{0} & i \boldsymbol{\sigma}^3 \end{pmatrix} \begin{pmatrix} \alpha \\ \beta \end{pmatrix} = \begin{pmatrix} i \boldsymbol{\sigma}^3 \alpha \\ i \boldsymbol{\sigma}^3 \beta \end{pmatrix}.
\end{equation}
If $\boldsymbol{\sigma}^3 \alpha = \alpha$ and $\boldsymbol{\sigma}^3 \beta = \beta$, then we have a candidate for $\mathcal{O}(i)$. The exact choice of $\alpha$ and $\beta$ will be determined by our additional requirement that $w$ be an eigenvector of $\gamma^0$:
\begin{equation}
\gamma^0 w = w.
\end{equation}
In the Dirac basis,
\begin{align*}
\begin{pmatrix} \mathbf{1} & \mathbf{0} \\ \mathbf{0} & -\mathbf{1} \end{pmatrix} \begin{pmatrix} \alpha \\ \beta \end{pmatrix} &=
\begin{pmatrix} \alpha \\ \beta \end{pmatrix} \\
\begin{pmatrix} \, \, \alpha \\ -\beta \end{pmatrix} &=
\begin{pmatrix} \alpha \\ \beta \end{pmatrix}.
\end{align*}
Therefore, $\alpha = u_+ \equiv \begin{pmatrix} 1 \\ 0 \end{pmatrix}$ and $\beta = 0 \equiv \begin{pmatrix} 0 \\ 0 \end{pmatrix}$, and the projection bispinor becomes
\begin{equation}
w = \begin{pmatrix} u_+ \\ 0 \end{pmatrix}.
\end{equation}
The bispinor now takes the form
\begin{align}
\psi & = \psi_M w \nonumber \\
& = \begin{pmatrix} f \, \mathbf{1} + i B_i \boldsymbol{\sigma}^i &  i g \, \mathbf{1} + E_i \boldsymbol{\sigma}^i \\ i g \, \mathbf{1} + E_i \boldsymbol{\sigma}^i & f \, \mathbf{1} + i B_i \boldsymbol{\sigma}^i \end{pmatrix}
\begin{pmatrix} u_+ \\ 0 \end{pmatrix} \nonumber \\
& = \begin{pmatrix} (f \, \mathbf{1} + i B_i \boldsymbol{\sigma}^i) u_+ \\ (i g \, \mathbf{1} + E_i \boldsymbol{\sigma}^i) u_+ \end{pmatrix} \nonumber \\
& = \begin{pmatrix} f + i B_3 \\ i B_1 - B_2 \\ i g + E_3 \\ E_1 + i E_2 \end{pmatrix}.
\end{align}

\section{Dirac Bilinears}
\label{sec:bilinears}

With the above choice of basis and projection bispinor, bilinears can easily be rewritten in terms of the multivector field components. The general procedure for constructing Dirac bilinears is as follows. The Dirac adjoint $\overline{\psi} = \psi^\dagger \gamma^0$ of the Dirac bispinor $\psi = \psi_M w$ is
\begin{align*}
\overline{\psi} & \equiv \psi^\dagger \gamma^0 = w^\dagger \psi_M^\dagger \gamma^0 \\
& = w^\dagger (\gamma^0 \widetilde{\psi}_M \gamma^0) \gamma^0 \\
& = w^\dagger \gamma^0 \widetilde{\psi}_M.
\end{align*}
For an operator $A$, the bilinear $\overline{\psi} A \psi$ now takes the form
\begin{equation*}
\overline{\psi} A \psi = w^\dagger \gamma^0 \widetilde{\psi}_M A \psi_M w \equiv w^\dagger M w.
\end{equation*}
With the chosen projection bispinor $w = \begin{pmatrix} u_+ \\ 0 \end{pmatrix}$, the product $w^\dagger M w$ singles out the $(1,1)$ component of the matrix $M$:
\begin{equation}
w^\dagger M w = M_{11}.
\end{equation}
In the Dirac basis, the only matrices with a $(1,1)$ component are $1$, $\gamma^0$, $I \gamma^3$, and $I \gamma^3 \gamma^0$. Taking the $(1,1)$ component of a matrix $M$ is therefore equivalent to isolating the terms proportional to these four basis elements:
\begin{equation}
M_{11} = \langle M \, 1 \rangle + \langle M \gamma^0 \rangle + i \, \langle M I \gamma^3 \rangle + i \, \langle M I \gamma^3 \gamma^0 \rangle.
\end{equation}
(In our matrix representation, extraction of the grade-0 terms can be accomplished by a trace operation: $\langle M \rangle = \frac{1}{4} \text{Tr} M$.)
 
Because all bilinears of physical interest are constructed to be real quantities, the imaginary $I \gamma^3$ and $I \gamma^3 \gamma^0$ terms will be zero, simplifying the expression to:
\begin{equation} \label{eqn:m11}
M_{11} = \langle M \, 1 \rangle + \langle M \gamma^0 \rangle.
\end{equation}
Furthermore, we are interested only in operators that correspond to physical quantities with well-defined transformation properties. The bilinears will therefore always be constructed so that $M$ is either purely even or purely odd, and only one of the terms in (\ref{eqn:m11}) will be non-zero.

\subsection{The scalar and pseudoscalar bilinears}
For the Lorentz scalar bilinear, we have
\begin{equation*}
\overline{\psi} \psi = w^\dagger \gamma^0 \widetilde{\psi}_M \psi_M w.
\end{equation*}
The operator $M = \gamma^0 \widetilde{\psi}_M \psi_M$ is the product of the (odd) vector $\gamma^0$ with two even multivectors. It is therefore odd, and the scalar $\langle M \rangle$ is necessarily zero, leaving only:
\begin{equation*}
w^\dagger M w = \langle M \gamma^0 \rangle = \langle \gamma^0 \widetilde{\psi}_M \psi_M \gamma^0 \rangle.
\end{equation*}
By the cyclic property of the scalar term,
\begin{align*}
\langle \gamma^0 \widetilde{\psi}_M \psi_M \gamma^0 \rangle & = \langle \psi_M \gamma^0 \gamma^0 \widetilde{\psi}_M \rangle \\
& = \langle \psi_M \widetilde{\psi}_M \rangle.
\end{align*}
In terms of the field components of $\psi_M$, the scalar becomes
\begin{align}
\overline{\psi} \psi & = \langle \psi_M \widetilde{\psi}_M \rangle \nonumber \\
& = \langle f^2 \, 1 - E_i E_j \, \gamma^i \gamma^0 \gamma^j \gamma^0 - B_i B_j \, I \gamma^i \gamma^0 I \gamma^j \gamma^0 + g^2 I^2 + \cdots \rangle \nonumber \\
& = f^2 - g^2 + \mathbf{B}^2 - \mathbf{E}^2.
\end{align}
Setting $f = g = 0$, we recover the familiar Lorentz scalar of classical electromagnetism:
\begin{equation*}
\mathbf{B}^2 - \mathbf{E}^2.
\end{equation*}

The pseudoscalar bilinear is
\begin{equation*}
- i \overline{\psi} \gamma^5 \psi = - \overline{\psi} I \psi = - w^\dagger \gamma^0 \widetilde{\psi}_M I \psi_M w.
\end{equation*}
Here $M = - \gamma^0 \widetilde{\psi}_M I \psi_M$ is again odd, so that $\langle M \rangle = 0$, and
\begin{align*}
\langle M \gamma^0 \rangle & = - \langle \gamma^0 \widetilde{\psi}_M I \psi_M \gamma^0 \rangle \\
& = - \langle I (\psi_M \widetilde{\psi}_M) \rangle \\
& = - I \cdot (\psi_M \widetilde{\psi}_M),
\end{align*}
which isolates the pseudoscalar term of $\psi_M \widetilde{\psi}_M$. In terms of the field components,
\begin{equation}
-i \overline{\psi} \gamma^5 \psi = 2 (f g - \mathbf{E} \cdot \mathbf{B}).
\end{equation}
When $f = g = 0$, we recover the familiar electromagnetic pseudoscalar:
\begin{equation*}
-2 \mathbf{E} \cdot \mathbf{B}.
\end{equation*}

\subsection{The vector and pseudovector bilinears}

Applying the same procedure to the vector current $j^\mu = \overline{\psi} \gamma^\mu \psi$, our bilinear is:
\begin{equation*}
\overline{\psi} \gamma^\mu \psi = w^\dagger \gamma^0 \widetilde{\psi}_M \gamma^\mu \psi_M w.
\end{equation*}
With $M = \gamma^0 \widetilde{\psi}_M \gamma^\mu \psi_M$ an even product, $\langle M \gamma^0 \rangle$ is identically zero, leaving
\begin{align*}
\overline{\psi} \gamma^\mu \psi & = \langle M \rangle = \langle \gamma^0 \widetilde{\psi}_M \gamma^\mu \psi_M \rangle \\
& = \langle \gamma^\mu \psi_M \gamma^0 \widetilde{\psi}_M  \rangle \\
& = \gamma^\mu \cdot ( \psi_M \gamma^0 \widetilde{\psi}_M ).
\end{align*}
In other words, this quantity projects the $\gamma^\mu$ component of the vector $j \equiv \psi_M \gamma^0 \widetilde{\psi}_M$. We can see that $j$ is a vector, because it is odd and positive under reversal:
\begin{equation*}
({\psi_M \gamma_0 \widetilde{\psi}_M})^\sim = \psi_M \gamma_0 \widetilde{\psi}_M.
\end{equation*}
Multiplying out the product in terms of the field components,
\begin{align}
j^0 & = f^2 + g^2 + \mathbf{E}^2 + \mathbf{B}^2, \nonumber \\
j^i & = 2 (f \mathbf{E} + g \mathbf{B} + \mathbf{E} \times \mathbf{B})_i.
\end{align}
When $f = g = 0$, we recover the familiar conserved energy and momentum densities of free-field electromagnetism:
\begin{align*}
j^0_{em} = \mathbf{E}^2 + \mathbf{B}^2, \\
j^i_{em} = 2 (\mathbf{E} \times \mathbf{B})_i.
\end{align*}

The pseudovector bilinear is:
\begin{align*}
j_5 & \equiv \overline{\psi} \gamma^5 \gamma^\mu \psi \\
& = \overline{\psi} (-i I) \gamma^\mu \psi \\
& = w^\dagger \gamma^0 \widetilde{\psi}_M I \gamma^\mu \psi_M \gamma^1 \gamma^2 w \\
& = \langle \gamma^0 \widetilde{\psi}_M I \gamma^\mu \psi_M \gamma^1 \gamma^2 \rangle \\
& = \langle I \gamma^\mu (\psi_M I \gamma^3 \widetilde{\psi}_M) \rangle \\
& = (I \gamma^\mu) \cdot (\psi_M I \gamma^3 \widetilde{\psi}_M),
\end{align*}
where $\psi_M I \gamma^3 \widetilde{\psi}_M$ is odd and negative under reversal, as expected for a pseudovector.
In component form:
\begin{align}
j^0_5 & = 2 (-f E_3 - g B_3 + E_1 B_2 - E_2 B_1), \nonumber \\
j^1_5 & = - 2 (g E_2 - f B_2 + B_3 B_1 + E_3 E_1), \nonumber \\
j^2_5 & = - 2 (- g E_1 + f B_1 + B_3 B_2 + E_3 E_2), \nonumber \\
j^3_5 & = - (f^2 + g^2 + B_3^2 - B_1^2 - B_2^2 + E_3^2 - E_1^2 - E_2^2).
\end{align}
With $f=g=0$, the corresponding components in free-field electromagnetism are:
\begin{align*}
j^0_{5,em} & = 2 (E_1 B_2 - E_2 B_1) = 2 S_3, \\
j^1_{5,em} & = - 2 (B_3 B_1 + E_3 E_1) = -2 \sigma_{3 1}, \\
j^2_{5,em} & = - 2 (B_3 B_2 + E_3 E_2) = -2 \sigma_{3 2}, \\
j^3_{5,em} & = - (B_3^2 - B_1^2 - B_2^2 + E_3^2 - E_1^2 - E_2^2) = -2 \sigma_{3 3},
\end{align*}
where $S_i$ and $\sigma_{i j}$ are the field momentum density and Maxwell stress tensor components, respectively.

\subsection{The Dirac Lagrangian}
The Lagrangian for the massless free Dirac spinor is:
\begin{align} \label{eqn:lagrangian}
\mathcal{L} & = \text{Re}(i \overline{\psi} \gamma^\mu \partial_\mu \psi) \nonumber \\
& = - \text{Re}(w^\dagger \gamma^0 \widetilde{\psi}_M \gamma^\mu \partial_\mu \psi_M \gamma^1 \gamma^2 w) \nonumber \\
& = - \langle \gamma^0 \widetilde{\psi}_M \gamma^\mu \partial_\mu \psi_M \gamma^1 \gamma^2 \rangle \nonumber \\
& = - \langle \gamma^\mu \partial_\mu \psi_M \gamma^1 \gamma^2 \gamma^0 \widetilde{\psi}_M \rangle \nonumber \\
& = - \langle \gamma^\mu \partial_\mu \psi_M I \gamma^3 \widetilde{\psi}_M \rangle \nonumber \\
& = B_3 \partial_t f - f \partial_t B_3 - (\vec{B} \times \partial_t \vec{B})_3 \nonumber \\
& - E_3 \partial_t g + g \partial_t E_3 - (\vec{E} \times \partial_t \vec{E})_3 \nonumber \\
& - f \partial_3 g + g \partial_3 f + B_3 (\vec{\nabla} \cdot \vec{E}) - E_3 (\vec{\nabla} \cdot \vec{B}) \nonumber \\
& - (\vec{B} \times \vec{\nabla} g)_3 - f (\vec{\nabla} \times \vec{E})_3 \nonumber \\
& - (\vec{B} \times (\vec{\nabla} \times \vec{E}))_3 + (\vec{E} \times (\vec{\nabla} \times \vec{B}))_3 \nonumber \\
& - (\vec{E} \times \vec{\nabla} f)_3 - g (\vec{\nabla} \times \vec{B})_3.
\end{align}
Applying the Euler-Lagrange equations, we can verify that we recover the general Maxwell equations (\ref{eqn:maxwell}), demonstrating that we have an alternative formulation of the Lagrangian for classical electromagnetism in terms of the electromagnetic tensor components $F^{\mu \nu}$ instead of the electromagnetic potentials $A^\mu$.

\subsection{The momentum density}
The (asymmetric) energy-momentum tensor $T^{\mu \nu} = \text{Re}(-i \overline{\psi} \gamma^\mu \partial^\nu \psi)$ yields the conserved momentum density
\begin{align}
P^i & = T^{0 i} = \text{Re}(-i \overline{\psi} \gamma^0 \partial^i \psi) \nonumber \\
& = \text{Re}(w^\dagger \gamma^0 \widetilde{\psi}_M \gamma^0 \partial^i \psi_M \gamma^1 \gamma^2 w) \nonumber \\
& = \langle \gamma^0 \widetilde{\psi}_M \gamma^0 \partial^i \psi_M \gamma^1 \gamma^2 \rangle \nonumber \\
& = \langle \gamma^0 \partial^i \psi_M \gamma^1 \gamma^2 \gamma^0 \widetilde{\psi}_M \rangle \nonumber \\
& = \langle \gamma^0 \partial^i \psi_M I \gamma^3 \widetilde{\psi}_M \rangle \nonumber \\
& = (\partial^i B_3) f - (\partial^i f) B_3 - (\partial^i B_1) B_2 + (\partial^i B_2) B_1 \nonumber \\
& - (\partial^i E_3) g + (\partial^i g) E_3 - (\partial^i E_1) E_2 + (\partial^i E_2) E_1.
\end{align}
The electromagnetic analogue, with $f = g = 0$, is
\begin{equation*}
P^i_{em} = - (\partial^i B_1) B_2 + (\partial^i B_2) B_1 - (\partial^i E_1) E_2 + (\partial^i E_2) E_1.
\end{equation*}
This is of course not the familiar electromagnetic field momentum density, which we've already seen in the guise of the current $j^i$, but another conserved ``momentum'' derived from our alternate Lagrangian (\ref{eqn:lagrangian}).

\subsection{The angular momentum density}
The conserved angular momentum density
\begin{equation*}
M^{\mu \nu \rho} = \text{Re}(-i \overline{\psi} \gamma^\mu (x^\nu \partial^\rho - x^\rho \partial^\nu) \psi + \frac{i}{4} \overline{\psi} \gamma^\mu \lbrack \gamma^\nu , \gamma^\rho \rbrack \psi)
\end{equation*}
yields an expression for the angular momentum in the $e_i - e_j$ plane (with $l^{ij} \equiv x^i \partial^j - x^j \partial^i$):
\begin{align*}
M^{0ij} & = \text{Re}(-i \overline{\psi} \gamma^0 l^{ij} \psi + \frac{i}{4} \overline{\psi} \gamma^0 \lbrack \gamma^i , \gamma^j \rbrack \psi) \\
& = \text{Re}(w^\dagger \gamma^0 \widetilde{\psi}_M \gamma^0 l^{ij} \psi_M \gamma^1 \gamma^2 w - \frac{1}{2} w^\dagger \gamma^0 \widetilde{\psi}_M \gamma^0 (\gamma^i \wedge \gamma^j) \psi_M \gamma^1 \gamma^2 w) \\
& = \langle \gamma^0 \widetilde{\psi}_M \gamma^0 l^{ij} \psi_M \gamma^1 \gamma^2 \rangle - \frac{1}{2} \langle \gamma^0 \widetilde{\psi}_M \gamma^0 (\gamma^i \wedge \gamma^j) \psi_M \gamma^1 \gamma^2 \rangle \\
& = \langle \gamma^0 l^{ij} \psi_M \gamma^1 \gamma^2 \gamma^0 \widetilde{\psi}_M \rangle - \frac{1}{2} \langle \gamma^0 (\gamma^i \wedge \gamma^j) \psi_M \gamma^1 \gamma^2 \gamma^0 \widetilde{\psi}_M \rangle \\
& = \langle \gamma^0 l^{i j} \psi_M I \gamma^3 \widetilde{\psi}_M \rangle - \frac{1}{2} \epsilon^{i j k} \langle (I \gamma^k) (\psi_M I \gamma^3 \widetilde{\psi}_M) \rangle.
\end{align*}
In terms of the field components, we have:
\begin{align}
M^{0 i j} & = (l^{ij} B_3) f - (l^{ij} f) B_3 \nonumber \\
& - (l^{ij} B_1) B_2 + (l^{ij} B_2) B_1 \nonumber \\
& - (l^{ij} E_3) g + (l^{ij} g) E_3 \nonumber \\
& - (l^{ij} E_1) E_2 + (l^{ij} E_2) E_1 \nonumber \\
& + \frac{1}{2} (f^2 + g^2 - \mathbf{E}^2 - \mathbf{B}^2 + 2 E^2_3 + 2 B^2_3).
\end{align}
The electromagnetic equivalent ($f = g = 0$) is
\begin{align*}
M^{0 i j} & = (l^{ij} B_1) B_2 + (l^{ij} B_2) B_1 \\
& - (l^{ij} E_1) E_2 + (l^{ij} E_2) E_1 \\
& + \frac{1}{2} (- \mathbf{E}^2 - \mathbf{B}^2 + 2 E^2_3 + 2 B^2_3).
\end{align*}
This angular momentum density for free-field electromagnetism has both an orbital and a spin-$1/2$ contribution. Half-integral spin appears therefore not to be inherently ``quantum'' in nature, but rather a consequence of the choice of Lagrangian.

\section{Lorentz transformations}
\label{sec:lorentz}

A multivector-valued field is manifestly Lorentz invariant under change of reference frame: each component transforms contravariantly to its corresponding basis element:
\begin{align*}
\psi_M & = f' \, 1 + \frac{1}{2!} F'^{\mu \nu} e'_\mu \wedge e'_\nu + \frac{1}{4!} g'^{\mu \nu \rho \sigma} e'_\mu \wedge e'_\nu \wedge e'_\rho \wedge e'_\sigma \\
& = f \, 1 + \frac{1}{2!} F^{\mu \nu} e_\mu \wedge e_\nu + \frac{1}{4!} g^{\mu \nu \rho \sigma} e_\mu \wedge e_\nu \wedge e_\rho \wedge e_\sigma.
\end{align*}
where the basis vectors transform as $e_\mu \to e'_\mu = R e_\mu \widetilde{R}$. To find the corresponding transformation rule for the field component $(\psi_M)_X$, we can contract $\psi_M$ with the basis element $X'$ in the transformed reference frame. For example, the bivector component $F'_{\mu \nu}$ is (up to a sign, depending on whether $e_\mu \wedge e_\nu$ squares to $+1$ or $-1$):
\begin{align*}
F'_{\mu \nu} & = \pm \langle (e'_\mu \wedge e'_\nu) \psi_M \rangle \\
& = \pm \langle (R (e_\mu \wedge e_\nu) \widetilde{R}) \psi_M \rangle \\
& = \pm \langle (e_\mu \wedge e_\nu) (\widetilde{R} \psi_M R) \rangle.
\end{align*}
Similar transformations hold for each component of the multivector, yielding a Lorentz transformation rule that takes the same form for all components:
\begin{equation}
( \psi_M )_X \to (\psi_M )_{X'} = (\psi_M)_{R X \widetilde{R}} = ( \widetilde{R} \psi_M R )_X.
\end{equation}
Rewriting the inverse transformation as $S \equiv \widetilde{R}$, we can factor the Lorentz transformation of the Dirac bispinor into the expected two-sided transformation for the multivector field components, and a one-sided transformation for the projection bispinor $w$:
\begin{align*}
\psi & \to S \psi \\
& = S \psi_M (\widetilde{S} S) w \\
& = (S \psi_M \widetilde{S}) S w.
\end{align*}
Although $w$ contains no field degrees of freedom, the transformation rule for $w$ serves an important role: it guarantees the proper transformation of the reference multivectors $\gamma^0$ and $\gamma^1 \gamma^2$ that appear when constructing bilinears. The eigenvector relation $w = \gamma^0 w$ transforms as:
\begin{align*}
w & \to S w \\
& = S (\gamma^0 w) \\
& = S \gamma^0 (\widetilde{S} S) w \\
& = (S \gamma^0 \widetilde{S}) S w,
\end{align*}
and the operation $i w = -\gamma^1 \gamma^2 w$ transforms as:
\begin{align*}
i w & \to i S w \\
& = S i w \\
& = - S \gamma^1 \gamma^2 w \\
& = - S \gamma^1 \gamma^2 (\widetilde{S} S) w \\
& = - (S \gamma^1 \gamma^2 \widetilde{S}) S w.
\end{align*}
Because both $\gamma^0$ and $\gamma^1 \gamma^2$ transform via the same transformation law as $\psi_M$, so will bilinears constructed from them. For example, the $j^\mu$ component of the vector current transforms as:
\begin{align*}
\overline{\psi} \gamma^\mu \psi & = w^\dagger \gamma^0 \widetilde{\psi}_M \gamma^\mu \psi_M w \\
& \to (w^\dagger S^\dagger) \gamma^0 (S \widetilde{\psi}_M \widetilde{S}) \gamma^\mu (S \psi_M \widetilde{S}) S w \\
& = w^\dagger (\gamma^0 \widetilde{S} \gamma^0) \gamma^0 (S \widetilde{\psi}_M \widetilde{S}) \gamma^\mu (S \psi_M \widetilde{S}) S w \\
& = \langle (\gamma^0 \widetilde{S}) (S \widetilde{\psi}_M \widetilde{S}) \gamma^\mu (S \psi_M \widetilde{S}) S \rangle \\
& = \langle \gamma^\mu (S \psi_M \widetilde{S}) (S \gamma^0 \widetilde{S}) (S \widetilde{\psi}_M \widetilde{S}) \rangle \\
& = \langle \gamma^\mu (S \psi_M \gamma^0 \widetilde{\psi}_M \widetilde{S}) \rangle \\
& = \gamma^\mu \cdot (S j \widetilde{S}) \\
& = j'^\mu.
\end{align*}

No elements of the Geometric Algebra have the one-sided passive transformation rule $w \to S w$ of the projection bispinor. All of our multivector fields and bilinears transform as $M \to S M \widetilde{S}$. In the final analysis, therefore, the bispinor serves no physical role here. The scalar, bivector, and pseudoscalar fields, either alone or in combination with fixed multivectors such as $\gamma^0$ and $\gamma^1 \gamma^2$, provide all necessary dynamic degrees of freedom.

\section{Summary}
We have a factorization $\psi = \psi_M w$ for the Dirac wavefunction, where the factor
\begin{equation*}
\psi_M = f \, 1 - E_i \, \gamma^i \gamma^0 - B_i \, I \gamma^i \gamma^0 + g I
\end{equation*}
contains scalar, bivector, and pseudoscalar dynamic field degrees of freedom. If $\psi$ obeys the massless wave equation, then $\psi_M$ obeys Maxwell's equations, with electric and magnetic current densities $j_e = - \nabla f$ and $j_m = - \nabla g$. The constant, uniform projection bispinor $w$ serves to project these degrees of freedom onto an abstract vector space when we require column vector solutions of the wave equation. The fields all transform under rotations and boosts as expected for classical fields: $M \to S M \widetilde{S}$. The one-sided transformation $w \to S w$ of the projection bispinor guarantees that fixed reference multivectors such as $\gamma^0$ and $\gamma^1 \gamma^2$, when used to construct bilinears such as $j = \psi_M \gamma^0 \widetilde{\psi}_M$ and $j_5 = \psi_M I \gamma^3 \widetilde{\psi}_M$, will also transform correctly.

The application of concepts from the Geometric Algebra to quantum theory is not new: the translation of the Dirac bispinor $\psi$ into an even multivector $\psi_M$, and the expressions derived in section \ref{sec:bilinears} for the bilinears in terms of $\psi_M$, are known (see, e.g., \cite{doran-lasenby}). In those applications, however, the multivector is interpreted not as a set of scalar, bivector, and pseudoscalar fields, but as a transformation operator $\psi_M = a \exp{B}$ (with scalar $a$ and bivector $B$) that rotates, boosts, and scales fixed spinors:
\begin{equation*}
w \to \psi = w' = \psi_M w
\end{equation*}
and spacetime vectors, e.g.,
\begin{equation*}
\gamma^0 \to v = \psi_M \gamma^0 \widetilde{\psi}_M.
\end{equation*}
We believe that our alternative interpretation provides a more satisfactory clarification of the role of the bispinor. The similarity between the first-order massless wave equation (\ref{eqn:maxwell}) and Maxwell's equations also allows us not only to understand more clearly the structure of the Dirac bilinears, but also to reveal the presence of ``quantum''-like conserved quantities in classical electromagnetism.

\bibliography{on-the-physical-interpretation-of-the-dirac-wavefunction}

\begin{thebibliography}{1}

\bibitem{hestenes-space-time-algebra}
David Hestenes.
\newblock {\em Space-Time Algebra}.
\newblock Birkh{\"a}user Basel, 2015.

\bibitem{hestenes-new-foundations}
David Hestenes.
\newblock {\em New Foundations for Classical Mechanics}.
\newblock Springer, 1987.

\bibitem{doran-lasenby}
Chris Doran and Anthony Lasenby.
\newblock {\em Geometric Algebra for Physicists}.
\newblock Cambridge University Press, 2003.

\end{thebibliography}
\bibliographystyle{unsrt}
\end{document}